\documentclass[10pt, conference]{IEEEtran}
\usepackage{cite}

\ifCLASSINFOpdf
\usepackage[pdftex]{graphicx}
\else
\fi
\usepackage[cmex10]{amsmath}
\usepackage{algorithmic}
\usepackage[hyphens]{url}

\usepackage{amsthm}
\newtheorem{theorem}{Theorem}
\usepackage{adjustbox}
\usepackage{algorithm}
\usepackage{color}
\usepackage{multirow}
\usepackage{setspace}
\IEEEoverridecommandlockouts

\begin{document}
\title{Fixed-PSNR Lossy Compression for Scientific Data}

\author{\IEEEauthorblockN{Dingwen Tao,\IEEEauthorrefmark{1}
Sheng Di,\IEEEauthorrefmark{2}
Xin Liang,\IEEEauthorrefmark{3}
Zizhong Chen,\IEEEauthorrefmark{3} and
Franck Cappello\IEEEauthorrefmark{2}\IEEEauthorrefmark{4}}
\IEEEauthorblockA{\IEEEauthorrefmark{1}The University of Alabama, AL, USA
}
\IEEEauthorblockA{\IEEEauthorrefmark{2}Argonne National Laboratory, IL, USA}
\IEEEauthorblockA{\IEEEauthorrefmark{2}
University of California,
Riverside, CA, USA}
\IEEEauthorblockA{\IEEEauthorrefmark{4}University of Illinois at Urbana-Champaign, IL, USA}
tao@cs.ua.edu, sdi1@anl.gov, xlian007@ucr.edu,\\
chen@cs.ucr.edu, cappello@mcs.anl.gov}

\maketitle
\begin{abstract}
\linespread{1.0}\selectfont
Error-controlled lossy compression has been studied for years because of extremely large volumes of data being produced by today's scientific simulations. None of existing lossy compressors, however, allow users to fix the peak signal-to-noise ratio (PSNR) during compression, although PSNR has been considered as one of the most significant indicators to assess compression quality. In this paper, we propose a novel technique providing a fixed-PSNR lossy compression for scientific data sets. We implement our proposed method based on the SZ lossy compression framework and release the code as an open-source toolkit. We evaluate our fixed-PSNR compressor on three real-world high-performance computing data sets. Experiments show that our solution has a  high accuracy in controlling PSNR, with an average deviation of 0.1 $\sim$ 5.0 dB on the tested data sets.  
\end{abstract}
\section{Introduction}
\label{sec:intro}

Today's high-performance computing (HPC) applications are producing extremely large amounts of data, such that data storage, transmission, and postanalysis are becoming a serious problem for scientific research. The problem comes from the limited storage capacity and I/O bandwidth of parallel file systems in production facilities. For example, the cosmology  Hardware/Hybrid Accelerated Cosmology Code (HACC) \cite{hacc} is a typical example of parallel executions facing this issue. 
According to the HACC researchers, the number of particles to simulate can be up to 3.5 trillion in today's simulations (and even more in the future), which leads to 60 PB of data to store; yet a system such as the Mira supercomputer [4] has only 26 PB of file system storage. 
For now, HACC researchers have to control the data size by a temporal decimation (i.e., dumping the snapshots every $k$ time steps in the simulation), degrading the consecutiveness of simulation in time dimension and losing important information unexpectedly. To this end, lossy compression has been studied for years \cite{isc17, bigdata17}.

Peak signal-to-noise ratio (PSNR) is one of the the most critical indicators used to assess the distortion of reconstructed data versus original data in lossy compression. Unlike other evaluation metrics such as maximum compression error, PSNR measures the overall data distortion, which is closely related to the visual quality \cite{guthe2001real}. Most of the existing lossy data compressors (such as SZ \cite{sz17} and ZFP \cite{zfp}), however, are executed based on either a specific absolute error bound or a pointwise relative error bound. To adjust lossy compression for a particular data set to reach a target PSNR (or mean squared error (MSE)), users have to run the lossy compressor  multiple times each with different error-bound settings, a tedious and time-consuming task especially when  masses of fields need to be compressed (e.g., CESM simulation involves 100+ fields \cite{cesm}). In this paper, we explore a novel technique providing fixed-PSNR lossy compression for scientific users. 

Th contributions of this paper are as follows.
\begin{itemize}
\item We explore an in-depth analysis to precisely predict the overall data distortion (such as MSE and PSNR) based on state-of-the-art lossy compressors.
\item We propose a novel fixed-PSNR method that can allow users to fix PSNR during the lossy compression based on our accurate prediction of PSNR.
\item We implement our proposed fixed-PSNR method based on the SZ lossy compression framework and release the code as an open-source tool under a BSD license.
\item We evaluate our implemented fixed-PSNR method using three real-world HPC data sets. Experimental results show that our proposed solution can precisely control the overall data distortion (i.e., PSNR) accurately during the compression with negligible overhead. In absolute terms, the average deviation between the user-specified PSNRs (before the compression) and the actual PSNRs (after the decompression) can be limited within 0.1 $\sim$ 5.0 dB on the tested data sets.
\end{itemize}

The rest of the paper is organized as follows. In Section \ref{sec:background}, we discuss state-of-the-art lossy compression for scientific data and traditional error-control strategies. In Section \ref{sec:norm-preserve}, we discuss how to estimate the overall data distortion for $l^2$-norm-preserving lossy compression. In section \ref{sec:design}, we formulate equations to estimate the $l^2$-norm-based data distortion and propose our new fixed-PSNR error-control approach. In Section \ref{sec:evaluation}, we present our experimental evaluation results. In section \ref{sec:conclusion}, we summarize our conclusions and discuss the future work.
\section{Background}
\label{sec:background}
In this section, we first discuss state-of-the-art lossy compression methods for scientific data and then describe the traditional error-control strategies of existing lossy compressors.

\subsection{Scientific Data Compression}
Compression techniques for HPC scientific data have been studied for years. 
The data compressors can be split into two categories: lossless and lossy. Lossless compressors make sure that the reconstructed data set after the decompression is exactly the same as the original data set. Such a constraint may significantly limit the compression ratio (up to 2 in general \cite{lossless2006}) on the compression of scientific data. The reason is that scientific data are composed mainly of floating-point values and their tailing mantissa bits could be too random to compress effectively. State-of-the-art lossy compressors include SZ \cite{sz16, sz17}, ZFP \cite{zfp}, ISABELA \cite{isabela}, FPZIP \cite{fpzip}, SSEM \cite{ssem}, and NUMARCK \cite{numarck}. Basically, they can be categorized into two models: prediction based and transform based. A \textit{prediction-based compressor} predicts data values for each data point and encodes the difference between every predicted value and its corresponding real value based on a quantization method. Typical examples are SZ \cite{sz16, sz17}, ISABELA \cite{isabela}, and FPZIP \cite{fpzip}. A \textit{transform-based compressor} transforms the original data to another space where most of the generated data is close to zero, such that the data can be stored with a certain loss in terms of user-required error bounds. For instance, SSEM \cite{ssem} and ZFP \cite{zfp} adopt a discrete Wavelet transform (DWT) and a customized orthogonal transform, respectively.

SZ \cite{sz16, sz17} is a state-of-the-art error-bounded lossy compressor designed for HPC scientific data. As discussed above, it falls into the category of prediction-based model and comprises \textit{three} main steps: (1) predicting the value for each data point by its preceding neighbors in the multidimensional space; (2) performing a customized Huffman coding \cite{huffman} to shrink the data size significantly; and (3) applying the GZIP lossless compressor \cite{gzip} on the encoded bytes to further improve the compression ratio. The current version (v 1.4) of SZ \cite{sz17} adopts the Lorenzo predictor \cite{lorenzo} as the default prediction method in Step (1).

\subsection{Traditional Error-Control Methods of Lossy Compression}
Existing state-of-the-art error-bounded lossy compressors provide a series of different error-control strategies to satisfy the demands of users. For example, ISABELA \cite{isabela} can guarantee the relative error of each reconstructed data point fall into a user-set pointwise relative error bound.
ZFP \cite{zfp} has three compression modes:  fixed-accuracy mode, fixed-rate mode, and fixed-precision mode. Specifically, the fixed-accuracy mode can ensure the absolute error of each reconstructed data point less than a user-set absolute error bound. The fixed-rate mode can guarantee that the data will be compressed to a fixed number of bits. The fixed-precision mode can ensure the number of retained bits per value, which can lead to a control of pointwise relative error.
SZ \cite{sz17}  offers three methods to control different types of errors: absolute error, pointwise relative error, and value-range-based relative error. Note that unlike the pointwise relative error that is compared with the value of each data point, value-range-based relative error is compared with the value range. 
To the best of our knowledge, this work is the first attempt to design an error-control approach that can precisely control the overall data distortion (such as MSE and PSNR) for lossy compression of scientific data.
\section{$L^2$-Norm-Preserving Lossy Compression}
\label{sec:norm-preserve}
In this section, we first introduce the prediction-based lossy compression. Then we infer that the pointwise compression error (i.e., the difference between the original value of each data point and its decompressed value) is equal to the error introduced by quantization \cite{quantization} or embedded coding (EC) \cite{ec}. We present two theorems to demonstrate that the overall distortion-based $l^2$ norm of the  reconstructed data can stay the same as the distortion in the second step. 

In the compression phase of the prediction-based lossy compression, the first step is to predict the value of each data point and calculate the prediction errors. 
This  step will generate a set of prediction errors (denoted by $X_{pe}$) based on the original data set (denoted by $X$), data point by data point, during the compression. 
The second step is to quantize or encode these prediction errors. The third step is optional for further reducing the data size by entropy encoding.

During the decompression, one needs to reconstruct the prediction errors based on quantization method or EC and then reconstruct the overall data set.
Note that the procedure of constructing the decompressed data set (denoted $\tilde{X}$), data point by data point, is based on the reconstructed prediction errors (denoted $\tilde{X}_{pe}$) during the decompression.

In what follows, we derive that the following equation must hold for prediction-based lossy compression.
\begin{equation}
\label{eq:x1}
X - \tilde{X} = X_{pe} - \tilde{X}_{pe}
\end{equation}


During the compression, the prediction method generally predicts the value of each data point based on the data points nearby in space because of the potential high consecutiveness of the data set. The Lorenzo predictor \cite{lorenzo}, for example, approximates each data point by the values of its preceding adjacent data points.
Since the neighboring data values to be used to reconstruct each data point during the decompression are actually the decompressed values instead of the original values, in practice, one has to assure that the compression and decompression stage have exactly the same prediction procedure (including the data values used in the prediction method); 
otherwise, the data loss will be propagated during the decompression. Hence, the predicted values during the compression must be equal to the predicted values during the decompression. That is, we have $X_{pred} = \tilde{X}_{pred}$. Then, we can derive Equation (\ref{eq:x1}) based on the following two equations: $X_{pe} = X - X_{pred}$ and $\tilde{X} = \tilde{X}_{pe} + \tilde{X}_{pred}$.

Based on Equation (\ref{eq:x1}), we can easily derive the following theorem.
\begin{theorem}
\label{thm:1}
For prediction-based lossy compression, the overall $l^2$-norm-based data distortion is the same as the distortion (introduced in the second step) of the prediction error.
\end{theorem}

Theorem \ref{thm:1} can be extended to the transform-based lossy compression methods with orthogonal transforms (such as DWT), as shown below. Because of space limitations, we omit the details of the proof.

\begin{theorem}
\label{thm:2}
For orthogonal-transform-based lossy compression, the overall $l^2$-norm-based data distortion is the same as the distortion (introduced in the second step) of the transformed data.
\end{theorem}

This theorems indicates that \textit{the overall distortion of the  decompressed data can be estimated by the data distortion introduced in the second step for $l^2$-norm-preserving lossy compression}.

\section{Design of Fixed-PSNR Lossy Compression}
\label{sec:design}
In this section, we propose our design of fixed-PSNR error-control method for lossy compression of scientific data.

As shown in Theorem \ref{thm:1} and \ref{thm:2}, the $l^2$-norm-based error, such as MSE, introduced by the second step of prediction-based and orthogonal-transform-based lossy compression stays unchanged after decompression. Therefore, we can estimate the $l^2$-norm based compression error by estimating the error of the second step. For the following discussion, we focus mainly on \textit{quantization} rather than EC for the second step.

Quantization converts the prediction errors or transformed data (denoted by $X^t$) to another set of integer values that are easier to compress. In quantization, the value range is split into multiple bins. Then, the compressor goes through all the prediction errors or transformed data to determine in which bins they are located, and it represents their values by the corresponding bin indexes, which are integer values.
During the decompression, the midpoint of each quantization bin will be used to reconstruct the data  located in the bin; it is called the quantized value in the following discussion.
The effectiveness of the data reduction in quantization depends on the distribution of the prediction errors or transformed data.

\begin{figure}
\centering
\includegraphics[scale=0.40]{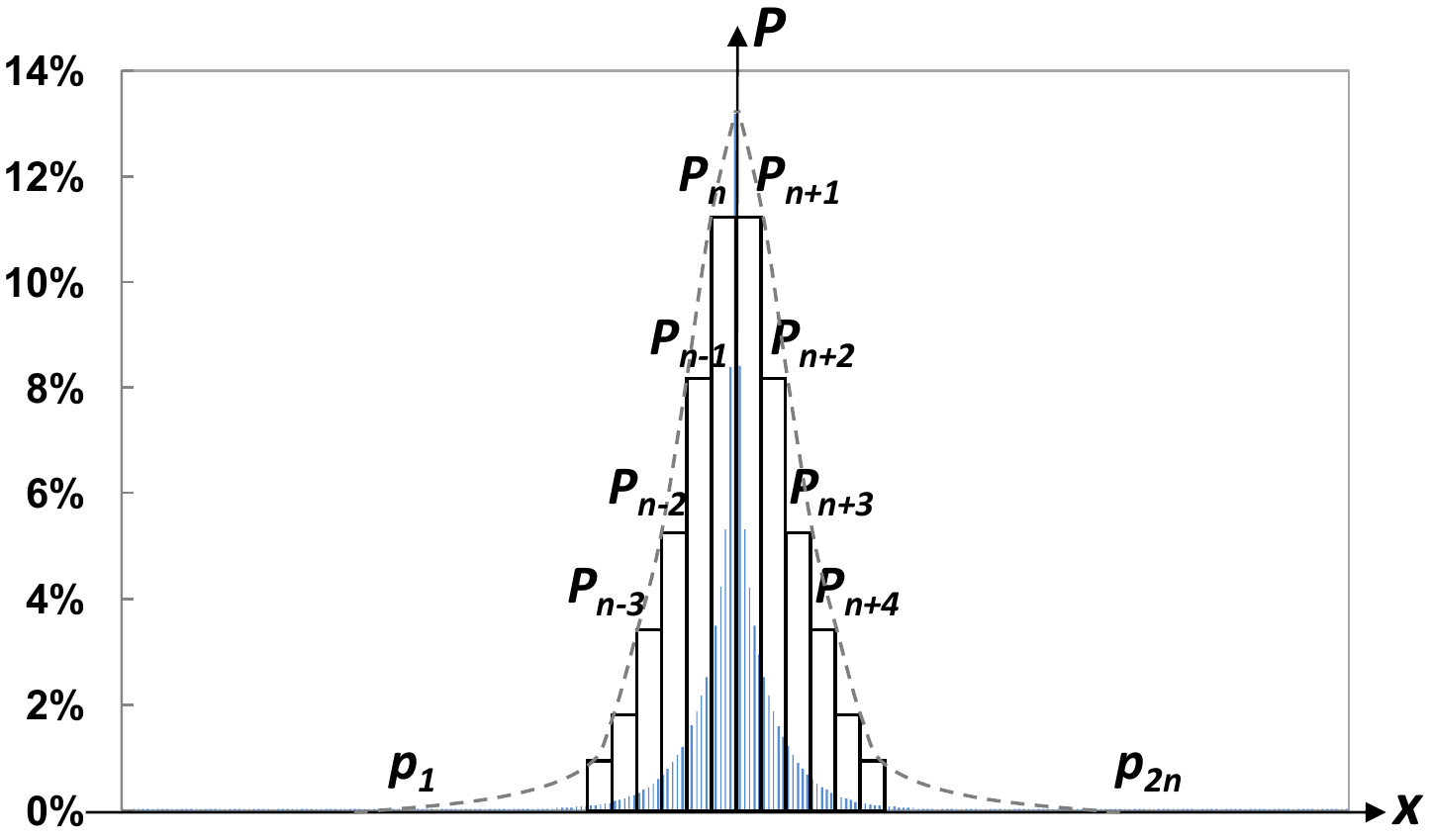}
\vspace{-2mm}
\caption{Example of the distribution and uniform quantization of the prediction errors generated by the SZ lossy compressor on one ATM data field.}
\label{fig:f1}
\vspace{-4mm}
\end{figure}

We assume $P(x)$ to be the probability density function of $X^t$, that is, $X^t \sim P(x)$.
The probability distribution of $X^t$ is symmetric in a large majority of cases, based on our experience. 
The blue area in Figure \ref{fig:f1} shows the typical probability distribution of the prediction errors generated by the SZ lossy compressor using a data field from a production climate simulation (called CSEM-ATM).
Therefore, we assume $P(x)$ to be symmetric without loss of generality, and we  focus only on $n$ bins on one side.
The number of quantization bins is represented by $2n$.
We denote $\delta_i$ the length of the $i$th quantization bin, where $\delta_i = \delta_{(2n+1)-i}$ ($1\leq i \leq n$) because of the symmetry property.

We denote $\tilde{X^t}$ as the quantized values of $X^t$.
The MSE between $X^t$ and $\tilde{X^t}$ can be calculated by
\begin{align}
& \resizebox{.55\hsize}{!}{$MSE(X^t,  \tilde{X^t}) = E_{X^t}[(X^t - \tilde{X^t})^2]$} \nonumber \\
& \resizebox{.45\hsize}{!}{$ = 2\int_{0}^{+\infty} (x - \tilde{x})^2 \cdot P(x) dx$}, \label{eq:mse-pdf}
\end{align}
where $E[\cdot]$ represents the expectation.
Note that $\tilde{x}$ is a step function, since the values in each bin are quantized to the same value.
Lossy compressors such as NUMARACK\cite{numarck}, SSEM\cite{ssem}, and SZ\cite{sz17} often use the midpoint of the quantization bin to approximate the values located in it.
Therefore,  $\tilde{x} = \frac{s_i+s_{i+1}}{2} = m_i$ when $ s_i \leq x < s_{i+1}$.
We can further estimate the MSE based on the probability density function $P(x)$ and the step function $\tilde{x}$ as follows.
\begin{align}
& \resizebox{.6\hsize}{!}{
$MSE = 2\sum\limits_{i=1}^{n} \int_{s_{i}}^{s_{i+1}} (x - \hat{x})^2 \cdot P(x) dx$} \nonumber \\
& \resizebox{.55\hsize}{!}{$\approx 2 \sum\limits_{i=1}^{n} (P(m_i) \cdot \int_{s_{i}}^{s_{i+1}} (x - m_i)^2 dx)$} \label{eq:approx} \\
& \resizebox{.75\hsize}{!}{$ = 2 \sum\limits_{i=1}^{n} (P(m_i) \cdot \int_{0}^{\delta_i} (x - \frac{\delta_i}{2})^2 dx) = \frac{1}{6} \sum\limits_{i=1}^{n} \delta_i^3 P(m_i)$} \nonumber
\end{align}

We then can calculate the normalized root mean squared error (NRMSE) and peak signal-to-noise ratio (PSNR) as follows:
\begin{align}
& \resizebox{.70\hsize}{!}{$NRMSE 	= \frac{\sqrt{MSE}}{vr} = (\sum\limits_{i=1}^{n} \delta_i^3  P(m_i))^{\frac{1}{2}} / \sqrt{6} vr$} \\
& \resizebox{.5\hsize}{!}{$PSNR 	= -20 \cdot \log_{10} (NRMSE)$} \nonumber \\
& \resizebox{.85\hsize}{!}{$ = -10 \cdot (\log_{10} (\sum\limits_{i=1}^{n} \delta_i^3  P(m_i)) + 2\cdot \log_{10} vr + \log_{10} 6)$},
\label{eq:psnr-pdf}
\end{align}
where $vr$ represents the value range of the original data $X$.
So far, we have established the estimation equation for $l^2$-norm-based compression error with general quantization. 

Uniform quantization is the most straightforward yet effective quantization approach, which is adopted by SZ lossy compressor.
With this approach, all quantization bins have the same length, (i.e.,  $\delta_1  = \cdots = \delta_{2n} = \delta$).
On the other hand, the $2n$ quantization bin can cover all the prediction errors as long as the number of bins is large enough; hence,
$\sum_{i=1}^{n} P(m_i)  \approx  P(x \geq 0)/\delta  = 1/2\delta$.
Equations (\ref{eq:psnr-pdf}) now can be further simplified as follows:
\begin{equation}
\label{eq:psnr-pdf-linear}
PSNR  = 20\cdot \log_{10} (vr/\delta) + 10\cdot \log_{10} 12.
\end{equation}
Equation (\ref{eq:psnr-pdf-linear}) says that \textit{the PSNR depends only on the unified quantization bin size regardless of the distribution of $X^t$}.
For example, the SZ lossy compressor sets the bin size $\delta$ to twice the absolute error bound (i.e., $eb_{abs}$) to ensure the maximum pointwise compression error within $eb_{abs}$.
Therefore, based on Equation (\ref{eq:psnr-pdf-linear}), our PSNR estimation for SZ lossy compressor becomes
\begin{equation}
\label{eq:psnr-sz}
PSNR = 20\cdot \log_{10} (vr/eb_{abs}) + 10\cdot \log_{10} 3.
\end{equation}
Note that $eb_{abs}/vr$ is the value-range-based relative error bound (denoted by $eb_{rel}$) defined by SZ. Thus we can estimate SZ's PSNR precisely based on the user-set $eb_{rel}$.


Similar to the definitions of fixed-rate and fixed-accuracy, we can design a fixed-PSNR mode for the SZ lossy compression framework.
The term ``fixed'' here means that the compression can be performed based on a specific given target (such as bit-rate or accuracy).
More generally, the following theorem can be drawn from our analysis.

\begin{theorem}
\label{thm:3}
The prediction-based and orthogonal-transform-based lossy compression methods with uniform quantization in the second step are fixed-PSNR regardless of the prediction errors or transformed data. In other words, the $l^2$-norm-based compression error can be estimated precisely by the unified quantization bin size and the value range of original data.
\end{theorem}

We can further simplify Equation (\ref{eq:psnr-sz}) to the following equation. 
\begin{equation}
\label{eq:psnr-fixed}
eb_{rel} = \sqrt{3} \cdot 10^{-\frac{PSNR}{20}}
\end{equation}
Based on Equation (\ref{eq:psnr-fixed}), we  propose our fixed-PSNR error-control approach based on the SZ compression framework. It consists of the following three steps: (1) get user-required PSNR, (2) estimate value-range-based relative error bound based on Equation \ref{eq:psnr-sz}, and (3) perform original SZ lossy compression with the estimated $eb_{rel}$.
Compared with the original compression workflow, the only computational overhead of our approach is the time to calculate the value-range-based relative error bound based on Equation (\ref{eq:psnr-fixed}) for each data field, which is negligible.
\section{Experimental Evaluation}
\label{sec:evaluation}

In this section, we evaluate our proposed fixed-PSRN lossy compression on three real-world HPC simulation data sets.

\begin{table}
\centering
\caption{Data Sets Used in Experimental Evaluation}
\label{tab:data}
\begin{adjustbox}{max width=\columnwidth}
\begin{tabular}{|c|c|c|c|c|}
\hline
                       & Dim. Size & \# of Fields & Data Size & Example Fields \\  \hline
\textbf{NYX}  & $2048 \times 2048 \times 2048$  & 6     & $206$ GB & baryon\_density, temperature \\
\hline
\textbf{ATM}          & $1800 \times 3600$  &  79    & $1.5$ TB & CLDHGH, CLDLOW\\ \hline
\textbf{Hurricane}  & $100 \times 500 \times 500$  & 13      & $62.4$ GB & QICE, PRECIP, U, V, W\\  \hline
\end{tabular}
\end{adjustbox}
\end{table}

We conduct our evaluation using the Bebop cluster \cite{bebop} at Argonne National Laboratory. Each node is equipped with two Intel Xeon E5-2695 v4 processors and 128 GB of memory. We perform our evaluations on various single floating-point HPC application data sets including 2D CESM-ATM data sets from climate simulation \cite{cesm}, 3D Hurricane data sets from the simulation of the hurricane ISABELA \cite{hurricane}, and 3D NYX data sets from cosmology simulation \cite{nyx}. Each application involves many simulation snapshots (or time steps). The details of the data sets are described in Table \ref{tab:data}.  

\begin{figure}
\centering
\includegraphics[scale=0.29]{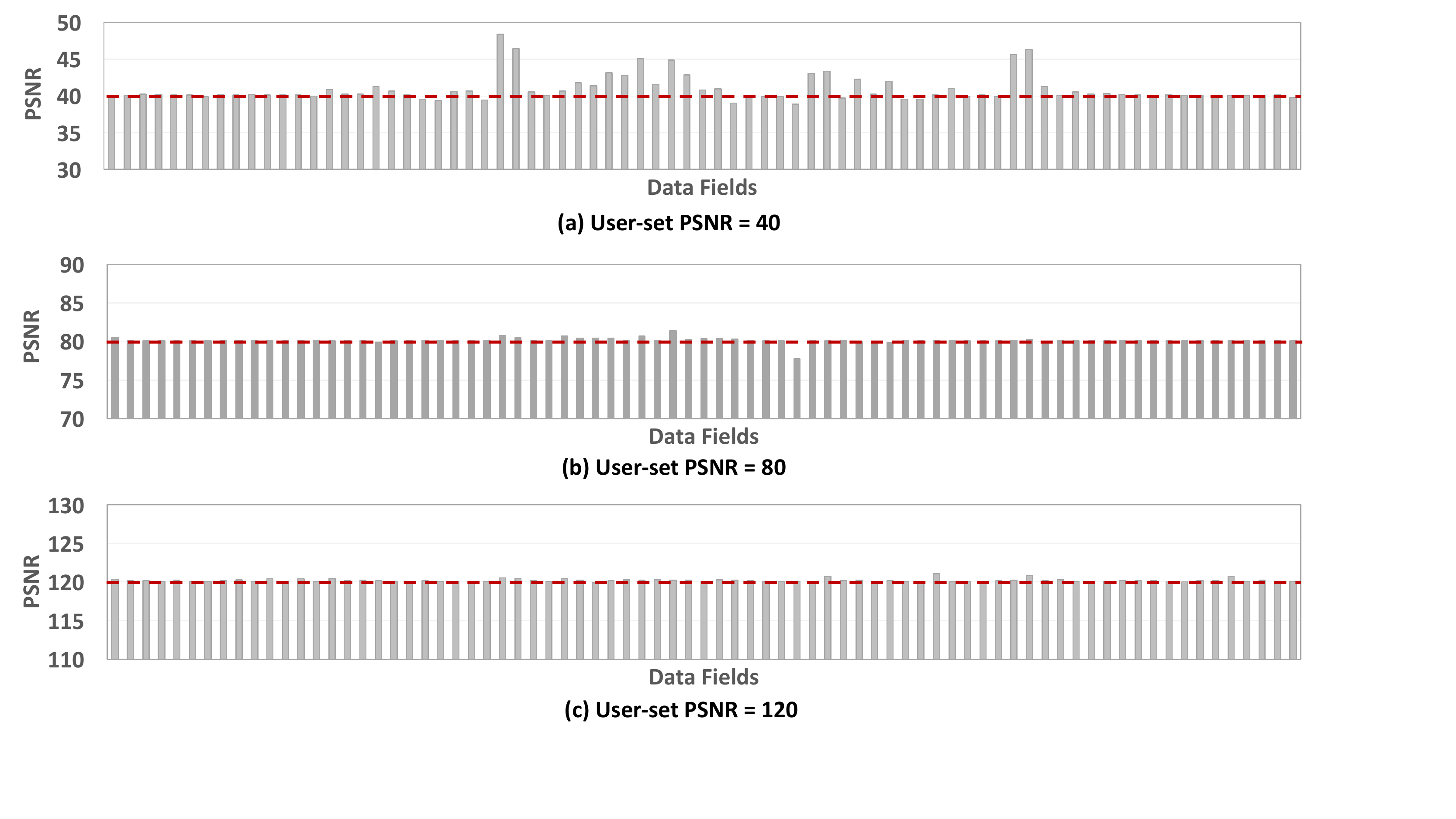}
\vspace{-1mm}
\caption{Evaluation of proposed fixed-PSNR mode on all data fields in ATM.}
\label{fig:f3}
\vspace{-2mm}
\end{figure}

\begin{table}
\centering
\caption{Evaluation of Proposed Fixed-PSNR Mode with SZ on NYX, ATM, and Hurricane Data Sets}
\vspace{-1mm}
\label{tab:results}
\begin{adjustbox}{max width=\columnwidth}
\begin{tabular}{|c|c|c|c|c|c|c|}
\hline
\multirow{2}{*}{\textbf{\begin{tabular}[c]{@{}c@{}}User-set\\ PSNR (dB)\end{tabular}}} & \multicolumn{2}{c|}{\textbf{NYX}} & \multicolumn{2}{c|}{\textbf{ATM}} & \multicolumn{2}{c|}{\textbf{Hurricane}} \\ \cline{2-7} 
                                                                                       & AVG             & STDEV           & AVG             & STDEV           & AVG                & STDEV              \\ \hline
 20                                                                                     & 24.3            & 1.82            & 21.9            & 3.34            & 25.0               & 6.52               \\ \hline
40                                                                                     & 41.9            & 2.32            & 40.9            & 1.80            & 42.0               & 3.97               \\ \hline
60                                                                                     & 60.7            & 0.74            & 60.2            & 0.62            & 60.5               & 0.74               \\ \hline
80                                                                                     & 80.1            & 0.05            & 80.1            & 0.35            & 80.1               & 0.32               \\ \hline
100                                                                                    & 100.1           & 0.07            & 100.2           & 0.17            & 100.1              & 0.39               \\ \hline
120                                                                                    & 120.1           & 0.01            & 120.2           & 0.19            & 120.3              & 0.63               \\ \hline
\end{tabular}
\end{adjustbox}
\vspace{-4mm}
\end{table}

Figure \ref{fig:f3} shows the evaluation results of our proposed fixed-PSNR lossy compression on all the fields in the ATM data sets with different user-set PSNRs. The three PSNR values---40 dB, 80 dB, and 120 dB---typically represent the low, medium, and high compression-quality scenarios, respectively. We plot the red dash lines showing the user-set PSNRs for ease comparison. Our fixed-PSNR lossy compressor can precisely control the PSNRs and meet the PSNR's demands for more than 90+\% ATM data fields on the average. Note that here ``meet'' means that the actual PSNR after the decompression is equal or higher than the user-set PSNR before the compression. 

Table \ref{tab:results} presents the experimental results of our proposed fixed-PSNR lossy compression on the NYX, ATM, and Hurricane data sets. The \textit{AVG} and \textit{STDEV} in the table represent the average value and standard deviation of the actual PSNRs of all the data fields in the set after the decompression.
The table illustrates that our solution provides a very high accuracy in controlling PSNR on these HPC data sets. Specifically, in absolute terms, the deviation between the user-set PSNRs and the actual PSNRs can be limited within 0.1 $\sim$ 5.0 dB on average.
We note that Figure \ref{fig:f3} and Table \ref{tab:results} both illustrate that the higher the PSNR of demand, the better our fixed-PSNR method performs. The reason can be that the accuracy of our estimation of the MSE (as shown in Equation \ref{eq:approx}) will decrease as the size of the quantization bin increases (i.e., higher error bound in the uniform quantization).
\section{Conclusion and Future Work}
\label{sec:conclusion}

In this paper, we propose a novel approach that can perform the lossy compression in terms of the user's specific demand on overall distortion of data (PSNR). To the best of our knowledge, this is the first attempt to control the overall data distortion, which provides  practical support to help efficiently tune the lossy compression to a target compression quality. Our experiments with three well-known real-world data sets show that the deviation between the user-specified PSNR before the compression and the actual PSNRs after the lossy compression can be limited within 0.1 $\sim$ 5.0 dB on the average. In future work, we will explore more techniques to further improve the fixed-PSNR lossy compression, especially for the low compression-quality demands. 

\section*{Acknowledge}
\footnotesize{This research was supported by the Exascale Computing Project (ECP), Project Number: 17-SC-20-SC, a collaborative effort of two DOE organizations – the Office of Science and the National Nuclear Security Administration, responsible for the planning and preparation of a capable exascale ecosystem, including software, applications, hardware, advanced system engineering and early testbed platforms, to support the nation’s exascale computing imperative. The material was supported by the U.S. Department of Energy, Office of Science, under contract DE-AC02-06CH11357, and supported by the National Science Foundation under Grant No. 1619253. We gratefully acknowledge the computing resources provided on Bebop, a high-performance computing cluster operated by the Laboratory Computing Resource Center at Argonne National Laboratory.}

\bibliographystyle{IEEEtran}
\bibliography{bib/refs}

\end{document}